\documentclass{article}
\usepackage{spconf,amsmath,graphicx}

\usepackage{booktabs}
\usepackage{multicol, multirow}
\usepackage{tabularx}
\usepackage{amssymb}
\usepackage{adjustbox}
\usepackage[dvipsnames]{xcolor}

\usepackage[ruled,vlined]{algorithm2e}

\title{de'HuBERT: Disentangling noise in a self-supervised model for robust speech recognition}
\name{\begin{tabular}{c} 
Dianwen Ng$^{1,2}$, Ruixi Zhang, Jia Qi Yip$^{1,2}$, Zhao Yang$^{2,3}$, Jinjie Ni$^{1,2}$, \\
Chong Zhang$^{1}$,
Yukun Ma$^{1}$
Chongjia Ni$^{1}$,
Eng Siong Chng$^{2}$, 
Bin Ma$^{1}$ \thanks{This work was supported by Alibaba Group through Alibaba Innovative
Research (AIR) Program and Alibaba-NTU Singapore
Joint Research Institute (JRI), Nanyang Technological University,
Singapore.}
\end{tabular}}

\address{$^1$Alibaba Group \\
 $^2$School of Computer Science and Engineering, Nanyang Technological University, Singapore \\
 $^3$Faculty of Electronic and Information Engineering, Xi’an Jiaotong University, China}
\begin{document}
%
\maketitle
\begin{abstract}
Existing self-supervised pre-trained speech models have offered an effective way to leverage massive unannotated corpora to build good automatic speech recognition (ASR). However, many current models are trained on a clean corpus from a single source, which tends to do poorly when noise is present during testing. Nonetheless, it is crucial to overcome the adverse influence of noise for real-world applications. In this work, we propose a novel training framework, called deHuBERT, for noise reduction encoding inspired by H. Barlow's redundancy-reduction principle. The new framework improves the HuBERT training algorithm by introducing auxiliary losses that drive the self- and cross-correlation matrix between pairwise noise-distorted embeddings towards identity matrix. This encourages the model to produce noise-agnostic speech representations. With this method, we report improved robustness in noisy environments, including unseen noises, without impairing the performance on the clean set. 
\end{abstract}
\section{Introduction}
\label{sec:intro}

Recently, self-supervised pre-training in speech has seized the limelight with numerous successes in building a highly effective automatic speech recognition (ASR) system \cite{baevski2020wav2vec, hsu2021hubert}, especially for low-resource languages \cite{babu2021xls}. This success stems from leveraging large amounts of unannotated utterances to construct universal speech representations that benefit downstream ASR tasks. Such frameworks include contrastive predictive coding (CPC) \cite{oord2018representation}, which learns by making the next step prediction using a contrastive loss, and autoregressive predictive coding (APC) \cite{chung2019unsupervised} that builds its speech representations by reconstructing future frames from the past sequence. 

Most of these works focused on a single domain of relatively clean audio, e.g. LibriSpeech \cite{panayotov2015librispeech}, that lacks domain variation. Nevertheless, speech in real-world environments usually contain background noises, reverberation and other non-linear distortions. \cite{hsu2021robust} had shown that many off-the-shelf universal speech models are vulnerable to this issue, where the performance of downstream ASR systems significantly degrade if there is a domain shift from the pre-training data.

To improve the noise robustness, \cite{wang2022wav2vec} modified wav2vec2.0 (w2v2) to include a contrastive loss that learns the cross-quantized targets between the original-noisy pair. Likewise, \cite{ng2022i2cr} employed contrastive loss as a regularizer to achieve noise-reduced speech features. \cite{huang2021spiral} provides another approach using a teacher-student framework to encode denoising representations from the perturbed data that resembles a siamese network. In addition, \cite{zhu2022noise} constructed an enhanced w2v2 that minimizes the consistency between noisy and clean features, and \cite{wang2022improving} introduced an auxiliary reconstruction task to improve the noise robustness of the learned representations. However, most of these approaches maybe hard to reproduce and involve careful implementation details.

In this paper, we aim to improve the noise robustness of the self-supervised pre-trained HuBERT \cite{hsu2021hubert} model for noisy ASR. We achieve this by introducing a new pair of auxiliary loss functions that encourages noise invariance in HuBERT's embedded contextual representations. To realize this, we propose a novel self-supervised training framework, disentangled HuBERT (deHuBERT), which regularizes HuBERT training using the recently proposed Barlow Twins \cite{zbontar2021barlow}, a method which reduces redundant information between the vector representations in images. We adapt this technique for sequential modelling and show that it is simple and highly effective in learning noise-invariant speech representations. The method aggregates the cross-correlation matrix between the embeddings of two identical networks forward-fed with different noise-augmented samples and pushes it towards the identity matrix. For the diagonal elements of the cross-correlation matrix to approach 1, the network has to extract agreeing features (i.e. speech content) of the two augmented utterances while minimizing other variational factors (i.e. background noises) between the dimensional representations at the frame level. Furthermore, decorrelating the off-diagonal elements creates the conditions for disentanglement. Experimental results show that our pre-trained model consistently exhibits better robustness in noisy environments, including unseen noises, without compromising the performance of the clean audio test set. 

\section{Methodology}
\label{sec:method}
\vspace{-0.1cm}
\subsection{HuBERT}
The HuBERT model architecture follows w2v2 with a convolutional encoder, BERT encoder, projection layer and code embedding layer. HuBERT adapts the BERT model from NLP to perform self-supervised speech representation learning. This allows the encoder to discover good high-level latent representations of both acoustic and language information from the continuous speech signals. During pre-training, it exploits an offline clustering step (i.e., using the K-Means algorithm) to generate the aligned discrete target labels (codes) for computing the BERT-like prediction loss from the masked frames, following the SpanBERT masking strategy. The training of HuBERT is initiated with hidden units of $K=100$ clusters derived based on the MFCC features of the raw audio data. In the subsequent iterations, the target codes are updated based on a hidden unit of $(K=500)$ clusters determined using the intermediate latent representations of the sixth layer of HuBERT's transformer at the second iteration. However, the HuBERT training algorithm does not inherently disentangle representations for noise separation or reduction, making the encoder vulnerable to noise. 

\subsection{deHuBERT}
\begin{figure}[t]
    \centering
    \includegraphics[width=1\linewidth, height=5cm]{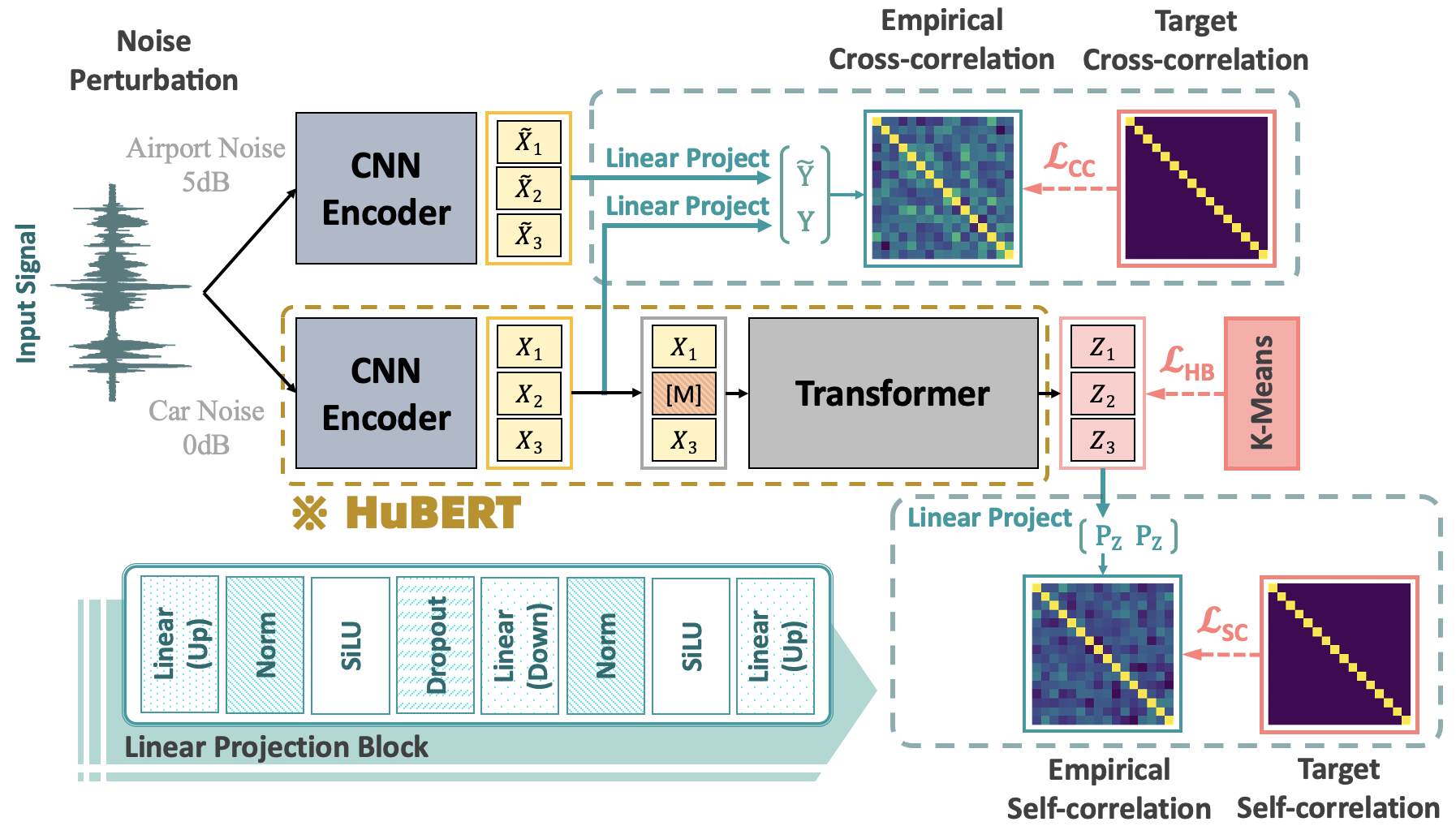}
    \caption{The deHuBERT method minimizes the self- and cross-correlation of the latent embeddings to an identity matrix.}
    \label{fig:proposedmodel}
    \vspace{-0.2cm}
\end{figure}

To obtain disentangled noise-agnostic representations using the HuBERT model, our proposed deHuBERT training algorithm makes use of the HuBERT to generate, in parallel, a second embedding of a different noise-augmented version using a shared CNN encoder, as shown in Fig \ref{fig:proposedmodel}. Here, two sets of noise are randomly selected and added to the training data with SNRs ranging between  0-25~dB. We then collect the encoded feature representations, $X$ and $\tilde{X}$, from the intermediate outputs and pass them to a shared linear projection block to get $Y$ and $\tilde{Y}$ respectively. Finally, following the losses introduced by ~\cite{zbontar2021barlow}, we derive the empirical cross-correlation (CC) matrix by
\begin{equation}
    C_{ij}^{(cc)} \triangleq \frac{\sum_n y_{n, i}\tilde{y}_{n, j} }{\sqrt{\sum_n (y_{n, i})^2} \sqrt{\sum_n (\tilde{y}_{n, j})^2}}
    \label{eqn:1}
\end{equation} where $n$ denotes the number of frames used and $i$, $j$ refer to the dimensional position of the frame-level representations. Note that $C \in [-1, 1]$ is a square matrix of $d$-dimensional based on the size of the projected output. We employ a CC loss that pushes the CC matrix towards the identity matrix. This loss function is defined by
\begin{equation}
    \mathcal{L}_{cc} \triangleq \underbrace{\sum_i (1 - C_{ii})^2}_{\text{invariance term}}  + \lambda \underbrace{\sum_i\sum_{j \neq i} {C_{ij}}^2}_{\text{disentangling term}}
    \label{eqn:2}
\end{equation} where $\lambda$ is a penalizing parameter that balances the trade off between the first and second terms of the loss.

Since $Y$ and $\tilde{Y}$ are sequential features, ignoring the frame-level correlation tends to overestimate the variability. To account for this, we flatten the outputs and remove the zero-padded frames within each minibatch before we perform a random sampling of size $n$, where we will index on both $Y$ and $\tilde{Y}$ identically. This causes the feature set to be more independent and will give us some control in tuning the stability of the proposed framework.

To understand how the proposed CC loss can reduce noise to obtain invariant features, we compare it to the infoNCE \cite{gutmann2010noise} loss. Formally, the first term in Eq. \ref{eqn:2} shares a close resemblance to the positive contrastive pair in infoNCE as presented in Eq. \ref{eqn:3}.
\begin{equation}
    \underbrace{-\sum_{n} \frac{\langle y_n, \tilde{y}_n\rangle_i}{\tau\|y_n\|_2 \|\tilde{y}_n\|_2}}_{\text{infoNCE's positive contrastive}} 
    \mspace{27mu}
    \underbrace{\sum_{i} (1 - \frac{\langle y_{.,i}, \tilde{y}_{.,i}\rangle_n}{\|y_{.,i}\|_2 \|\tilde{y}_{.,i}\|_2})^2}_{\text{proposed invariance term}}
    \label{eqn:3}
\end{equation} 
Similar to the objective behind the positive contrastive loss, we try to maximize the agreeing speech content between the two distorted embeddings and lower other variations (e.g. noise) by getting the two dimensional feature components perfectly correlated. Likewise, decorrelating the off-diagonal matrix discourages information sharing over the feature components while simultaneously encouraging disentangled representations. 

To gain further disentanglement in the output representations, we build another linear projection block of the same structure that takes in the bottleneck representations $Z$ to compute the projected $P_Z$ for estimating the empirical self-correlation (SC). The estimation can be done by reusing the computational function in Eq. \ref{eqn:1} with random sampling, and replacing the arguments with ($P_Z$, $P_Z$). Again, we compute the SC loss similar to Eq. \ref{eqn:2} but with the SC matrix. In practice, we believe that CC loss may not be perfect in obtaining noise-invariant representations. Disentangling the bottleneck features and then using them to predict the hidden units (i.e. Hubert's codes) of the original clean training audio guides the encoder to detect the residual noise information and eventually suppressing it in the final contextual representations.  

The complete optimization loss used in our pre-training framework is given by
\begin{equation}
    \mathcal{L} = \mathcal{L}_{\text{HB}} + \alpha \mathcal{L}_{\text{CC}} + \beta \mathcal{L}_{\text{SC}}
\end{equation} 
where the three terms refer to the HuBERT loss, cross-correlation loss and self-correlation losses, respectively. $\alpha$ and $\beta$ have both been set to 0.5 in this work.

\section{Experiment}
\subsection{Data Description} We set up our data environments following \cite{prasad2021investigation, zhu2022noise} for performance comparisons. In our experiments, we use the full 960h of Librispeech for pre-training and the dev-clean corpus for the validation set. The noise dataset used for training is obtained from FreeSound \cite{font2013freesound}, which consists of 16kHz noise data which can be categorized into stationary (Type A) and non-stationary (Type B). The type A noises available are Car, Metro and Traffic noises. In the Type B category, Babble, Airport/Station, Cafe and AC/Vacuum noises are available. Each type of noise has 10 and 8 different audio streams in the training and testing sets, respectively. The total duration of the noise data is around 2h. During testing, 120 randomly chosen sub-files from the test-clean set of Librispeech are used, as per the standard procedure for testing on this dataset. In addition, LibriSpeech comes with pre-mixed noises at different SNRs between 0-20~dB, which ultimately makes up 4200 instances of noisy test data. The noise data and noisy test sets can be downloaded from the website\footnote{ https://github.com/archiki/Robust-E2E-ASR}. 

\subsection{Model Pre-training} We perform continual pre-training by utilizing the weights provided by the Fairseq toolkit for 250k steps. In our implementation, we construct the final projection block with the corresponding $d$-dimensional size of 2048 and 4096 for CC and SC. In contrast to \cite{zbontar2021barlow}, we observed a concave plot of the performance with the effect on increasing dimensionality of the projector network. Additionally, we sampled n=640, and we found that adopting a smaller sample size benefits early-stage learning as it contains a slightly higher estimation error that excites the network and allows the model to escape from the local minimum. However, this requires a smaller $\lambda=0.005$ to limit the adversity contributed by the estimation error. Finally, we also found that applying a smaller learning rate of 7e-5 leads to better model pre-training.

\subsection{Model Fine-tuning}
We used the best checkpoint from the pre-training and followed the typical base setup for 100h, 10h, 1h and 10m. The ASR finetuning involves only the HuBERT component. Additionally, we employed multi-conditioning training with the training noise of 0 to 20~dB. Finally, we tested our performance with the best checkpoint according to the validation WER for final evaluations.  

\section{Experimental Results}
\begin{table*}[bht]\centering\footnotesize
\tabcolsep=0.32cm
\renewcommand{\arraystretch}{1}
\caption{Experimental results on the given synthesized noisy data for various noise types of SNRs (0-20)dB without a LM.}
\vspace{0.1cm}
\begin{tabular}{lcccccccccc}
\hline
\multicolumn{1}{l|}{\multirow{3}{*}{Methods}} &
  \multicolumn{1}{c|}{\multirow{3}{*}{Pre-train}} &
  \multicolumn{9}{c}{WER (\%) under noisy (0 -- 20~dB) SNR and clean environment $\downarrow$} \\ \cline{3-11} 
\multicolumn{1}{l|}{} &
  \multicolumn{1}{c|}{} &
  \multicolumn{4}{c|}{Type-B noise} &
  \multicolumn{3}{c|}{Type-A noise} &
  \multicolumn{1}{c|}{\multirow{3}{*}{\begin{tabular}[c]{@{}c@{}c@{}} \\ Avg.\\ (noisy)\end{tabular}}} &
  {\multirow{3}{*}{\begin{tabular}[c]{@{}c@{}c@{}} \\ Clean\\ (subset)\end{tabular}}} \\ \cline{3-9}
\multicolumn{1}{l|}{} &
  \multicolumn{1}{c|}{} &
  \multicolumn{1}{c|}{Babble} &
  \multicolumn{1}{c|}{\begin{tabular}[c]{@{}c@{}}Airport/\\ Station\end{tabular}} &
  \multicolumn{1}{c|}{\begin{tabular}[c]{@{}c@{}}AC/\\ Vacuum\end{tabular}} &
  \multicolumn{1}{c|}{Cafe} &
  \multicolumn{1}{c|}{Traffic} &
  \multicolumn{1}{c|}{Metro} &
  \multicolumn{1}{c|}{Car} &
  \multicolumn{1}{c|}{} &
   \\ \hline
\multicolumn{11}{c}{Fine-tuning: 10-hours labeled (with additive FreeSound noise)} \\ \hline
\multicolumn{1}{l|}{HuBERT Base} &
  \multicolumn{1}{c|}{Clean} &
  33.71 &
  26.85 &
  23.82 &
  \multicolumn{1}{c|}{20.19} &
  19.05 &
  18.26 &
  \multicolumn{1}{c|}{12.91} &
  \multicolumn{1}{c|}{22.11} &
  13.5 \\
\multicolumn{1}{l|}{HuBERT Base} &
  \multicolumn{1}{c|}{FreeSound} &
  27.93 &
  22.33 &
  20.77 &
  \multicolumn{1}{c|}{17.58} &
  17.08 &
  17.30 &
  \multicolumn{1}{c|}{13.05} &
  \multicolumn{1}{c|}{19.43} &
  13.7 \\
\multicolumn{1}{l|}{deHuBERT (Ours)} &
  \multicolumn{1}{c|}{FreeSound} &
  \textbf{26.58} &
  \textbf{21.23} &
  \textbf{20.14} &
  \multicolumn{1}{c|}{\textbf{16.83}} &
  \textbf{16.05} &
  \textbf{15.74} &
  \multicolumn{1}{c|}{\textbf{11.95}} &
  \multicolumn{1}{c|}{\textbf{18.36}} &
  \textbf{12.8} \\ \hline
\multicolumn{11}{c}{Fine-tuning: 1-hour labeled (with additive FreeSound noise)} \\ \hline
\multicolumn{1}{l|}{HuBERT Base} &
  \multicolumn{1}{c|}{Clean} &
  49.72 &
  41.86 &
  39.98 &
  \multicolumn{1}{c|}{35.79} &
  34.42 &
  33.08 &
  \multicolumn{1}{c|}{26.74} &
  \multicolumn{1}{c|}{37.37} &
  \textbf{27.8} \\
\multicolumn{1}{l|}{HuBERT Base} &
  \multicolumn{1}{c|}{FreeSound} &
  42.54 &
  36.83 &
  36.11 &
  \multicolumn{1}{c|}{32.82} &
  32.19 &
  31.77 &
  \multicolumn{1}{c|}{27.60} &
  \multicolumn{1}{c|}{34.27} &
  29.1 \\
\multicolumn{1}{l|}{deHuBERT (Ours)} &
  \multicolumn{1}{c|}{FreeSound} &
  \textbf{41.74} &
  \textbf{36.27} &
  \textbf{35.54} &
  \multicolumn{1}{c|}{\textbf{32.41}} &
  \textbf{31.51} &
  \textbf{31.24} &
  \multicolumn{1}{c|}{\textbf{26.68}} &
  \multicolumn{1}{c|}{\textbf{33.63}} &
  28.4 \\ \hline
\multicolumn{11}{c}{Fine-tuning: 10-mins labeled (with additive FreeSound noise)} \\ \hline
\multicolumn{1}{l|}{HuBERT Base} &
  \multicolumn{1}{c|}{Clean} &
  70.25 &
  63.62 &
  61.89 &
  \multicolumn{1}{c|}{57.68} &
  55.41 &
  54.66 &
  \multicolumn{1}{c|}{47.95} &
  \multicolumn{1}{c|}{58.78} &
  48.4 \\
\multicolumn{1}{l|}{HuBERT Base} &
  \multicolumn{1}{c|}{FreeSound} &
  60.53 &
  56.31 &
  56.00 &
  \multicolumn{1}{c|}{52.92} &
  53.16 &
  52.58 &
  \multicolumn{1}{c|}{49.56} &
  \multicolumn{1}{c|}{54.44} &
  50.7 \\
\multicolumn{1}{l|}{deHuBERT (Ours)} &
  \multicolumn{1}{c|}{FreeSound} &
  \textbf{58.59} &
  \textbf{53.82} &
  \textbf{53.88} &
  \multicolumn{1}{c|}{\textbf{50.66}} &
  \textbf{49.67} &
  \textbf{49.71} &
  \multicolumn{1}{c|}{\textbf{45.80}} &
  \multicolumn{1}{c|}{\textbf{51.73}} &
  \textbf{47.1} \\ \hline \hline
\multicolumn{11}{c}{Fine-tuning: 100-hours labeled (with additive FreeSound noise)} \\ \hline
\multicolumn{1}{l|}{DEMUCS \cite{prasad2021investigation}} &
  \multicolumn{1}{c|}{FreeSound} &
  45.56 &
  36.98 &
  38.20 &
  \multicolumn{1}{c|}{27.02} &
  26.46 &
  23.22 &
  \multicolumn{1}{c|}{16.02} &
  \multicolumn{1}{c|}{30.49} &
  10.9 \\
\multicolumn{1}{l|}{AvT \cite{prasad2021investigation}} &
  \multicolumn{1}{c|}{No} &
  43.42 &
  35.32 &
  36.62 &
  \multicolumn{1}{c|}{27.06} &
  27.88 &
  24.28 &
  \multicolumn{1}{c|}{17.76} &
  \multicolumn{1}{c|}{30.33} &
  13.1 \\
\multicolumn{1}{l|}{Wav2vec 2.0 \cite{zhu2022noise}} &
  \multicolumn{1}{c|}{Clean} &
  47.50 &
  39.68 &
  38.84 &
  \multicolumn{1}{c|}{31.14} &
  29.22 &
  27.44 &
  \multicolumn{1}{c|}{18.24} &
  \multicolumn{1}{c|}{33.15} &
  14.0 \\
\multicolumn{1}{l|}{Wav2vec 2.0 \cite{zhu2022noise}} &
  \multicolumn{1}{c|}{FreeSound} &
  39.56 &
  32.50 &
  34.94 &
  \multicolumn{1}{c|}{25.22} &
  24.52 &
  22.48 &
  \multicolumn{1}{c|}{16.24} &
  \multicolumn{1}{c|}{27.92} &
  13.5 \\
\multicolumn{1}{l|}{EW2 \cite{zhu2022noise}} &
  \multicolumn{1}{c|}{FreeSound} &
  33.88 &
  27.36 &
  27.94 &
  \multicolumn{1}{c|}{22.08} &
  20.94 &
  19.84 &
  \multicolumn{1}{c|}{14.88} &
  \multicolumn{1}{c|}{23.85} &
  12.3 \\
\multicolumn{1}{l|}{HuBERT Base} &
  \multicolumn{1}{c|}{FreeSound} &
  22.52 &
  16.91 &
  15.94 &
  \multicolumn{1}{c|}{12.79} &
  12.43 &
  12.20 &
  \multicolumn{1}{c|}{8.39} &
  \multicolumn{1}{c|}{14.45} &
  9.4 \\
\multicolumn{1}{l|}{deHuBERT (Ours)} &
  \multicolumn{1}{c|}{FreeSound} &
  \textbf{21.25} &
  \textbf{16.02} &
  \textbf{14.93} &
  \multicolumn{1}{c|}{\textbf{11.94}} &
  \textbf{11.66} &
  \textbf{11.21} &
  \multicolumn{1}{c|}{\textbf{7.62}} &
  \multicolumn{1}{c|}{\textbf{13.52}} &
  \textbf{8.6} \\  \hline
\end{tabular}
\label{tbl:1}
\vspace{-0.4cm}
\end{table*}

\begin{figure}
    \centering
    \includegraphics[width=0.99\linewidth, height=3.8cm]{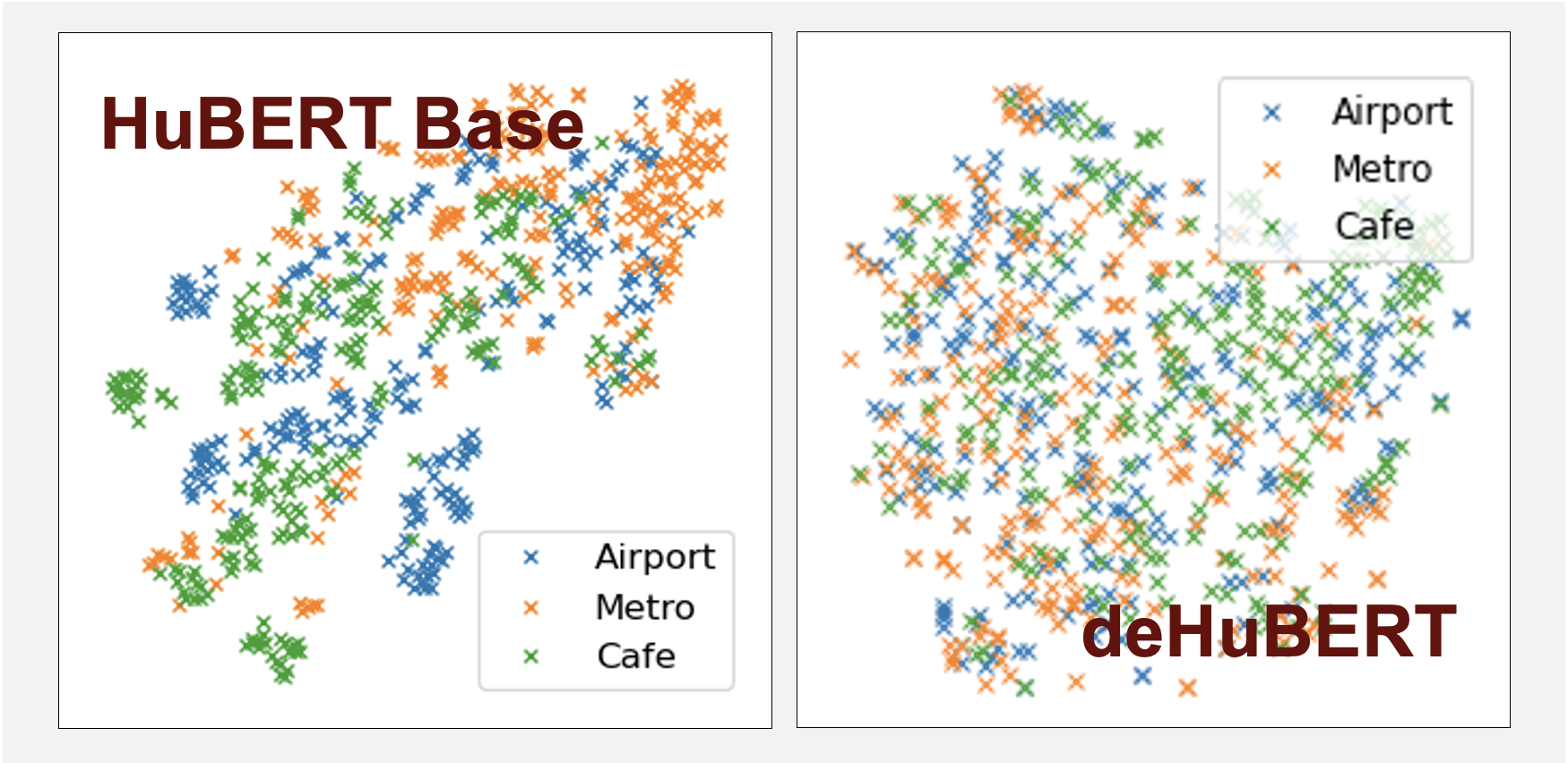}
    \vspace{-0.2cm}
    \caption{t-SNE plots that compare the disentanglement and noise invariance of different networks with 0~dB noises.}
    \label{fig:tsne}
    \vspace{-0.2cm}
\end{figure}

We compare our results without a language model with an off-the-shelf HuBERT as the baseline to determine the efficiency of our model in learning a noise-robust ASR with limited finetuning data. Also, we included results from the HuBERT base model that undergoes multi-conditioning pre-training to cast a holistic analysis. Table \ref{tbl:1} shows the ASR performance in WER based on the subset test-clean audio pre-mixed with the individual noise types of SNRs between 0 and 20~dB. We observe that pre-training HuBERT with noise helps to improve the adaptability to noise on the downstream ASR, but this comes at the cost of degrading clean speech performance. Nonetheless, deHuBERT outperforms baseline HuBERT on both noisy and clean speech regardless of the pre-training condition. Additionally, the difference in performance becomes more apparent with the increasing scarcity of finetuning resources. Finally, we investigate the experiment with the typical 100h finetuning to compare our deHuBERT with existing models. On the complete test-clean and test-other set, we achieved a WER of 6.3\% and 13.2\%, respectively. This score is comparable to the baseline performance despite using only noisy speech for finetuning. Additionally, deHuBERT achieves the top WER on the noisy data. 


To visualize the noise-agnostic properties of the deHuBERT embeddings, we plot the t-SNE of the bottleneck features of both HuBERT and deHuBERT in Fig. \ref{fig:tsne}. The features were obtained from 720 randomly selected audio samples of train-clean-100 mixed with 0~dB of Airport, Metro and Cafe noises. Before plotting, we performed a global mean pooling of all the bottle neck features in a sequence to get vector representations before applying the t-SNE algorithm. On the HuBERT Base plot (left), we can identify clusters consisting of samples with the same noise type, indicating the presence of noise information. In comparison, the deHuBERT plot (right) exhibits no clear clustering according to the type of noise. 

\vspace{-0.17cm}
\subsection{Post-methodology Study}
In this section, we are stress testing our model to determine the robustness of its out-of-domain (OOD) performance. We use the TEDLIUM3 \cite{hernandez2018ted} dataset to explore the effect of domain shift with noisy ASR. Moreover, we introduce out-of-domain office noise from FSD50K \cite{fonseca2021fsd50k} by selecting noise from the group \textit{Whispering, Writing, Typing, Typewriter, Telephone, Conversation, Laughter, Computer Keyboard} and \textit{Printer}. We filter those that are less than 10m, which led us to 385 files. Table \ref{tbl:2} presents the performance based on finetuning the selected clean audio set (10h) on the complete test set under three different conditions: (1) In-domain (ID) clean test set, (2) ID pre-train noise but OOD finetuning, (3) OOD pre-train noise and OOD finetuning. Firstly, our pre-trained model is comparable to the base under the condition (1). This is important as it indicates that our model remains robust and is unaffected by noisy pre-training. Secondly, even on unseen noise during finetuning, deHuBERT performs consistently better than HuBERT base under noisy environments in conditions (2) and (3). Lastly, although there is still a degradation in performance on ID and OOD noisy ASR, the percentage increase in WER is relatively lower in deHuBERT than for HuBERT base, especially for condition (3).  

\begin{table}[ht]\vspace{-0.1cm}
\centering\footnotesize
\renewcommand{\arraystretch}{1.1}
\caption{Results on various out-of-domain noisy conditions. We finetuned our model with 10h (respective) dataset.}
\vspace{0.15cm}
\begin{tabular}{l|c|cccc}
\hline
\multicolumn{1}{l|}{\multirow{3}{*}{Models}} &
  \multicolumn{1}{c|}{\multirow{3}{*}{\begin{tabular}[c]{@{}c@{}}FT Data\\ (10hrs)\end{tabular}}} &
  \multicolumn{4}{c}{WER (\%) of testing data $\downarrow$} \\ \cline{3-6} 
\multicolumn{1}{l|}{}                & \multicolumn{1}{c|}{}            & \multicolumn{2}{c|}{LS (Test set)} & \multicolumn{2}{c}{TEDLIUM} \\ \cline{3-6} 
\multicolumn{1}{l|}{} &
  \multicolumn{1}{c|}{} &
  \multicolumn{1}{c|}{Clean} &
  \multicolumn{1}{c|}{Other} &
  \multicolumn{1}{c|}{Dev} &
  Test \\ \hline
\multicolumn{6}{c}{Testing set from the original data (Clean)}                                                                       \\ \hline
\multicolumn{1}{l|}{HuBERT Base}     & \multicolumn{1}{c|}{LibriSpeech} & \textbf{9.8}   & \multicolumn{1}{c|}{18.2}   & \textbf{25.4}            & \textbf{23.6}           \\
\multicolumn{1}{l|}{deHuBERT (Ours)} & \multicolumn{1}{c|}{LibriSpeech} & 10.1   & \multicolumn{1}{c|}{\textbf{18.1}}   & 25.5            & 23.8            \\ \hline
\multicolumn{1}{l|}{HuBERT Base}     & \multicolumn{1}{c|}{TEDLIUM}     & \textbf{14.9}   & \multicolumn{1}{c|}{23.8}   & \textbf{18.1}            & \textbf{17.3}            \\
\multicolumn{1}{l|}{deHuBERT (Ours)} & \multicolumn{1}{c|}{TEDLIUM}     & 15.2   & \multicolumn{1}{c|}{\textbf{23.7}}   & 18.2            & 17.4            \\ \hline
\multicolumn{6}{c}{Testing set with additive FreeSound noise (0--20 dB)}                                                                       \\ \hline
\multicolumn{1}{l|}{HuBERT Base}     & \multicolumn{1}{c|}{LibriSpeech} & 20.3   & \multicolumn{1}{c|}{36.4}   & 35.8            & 36.4            \\
\multicolumn{1}{l|}{deHuBERT (Ours)} & \multicolumn{1}{c|}{LibriSpeech} & \textbf{13.4}   & \multicolumn{1}{c|}{\textbf{26.0}}   & \textbf{30.1}            & \textbf{30.3}            \\ \hline
\multicolumn{1}{l|}{HuBERT Base}     & \multicolumn{1}{c|}{TEDLIUM}     & 23.5   & \multicolumn{1}{c|}{38.8}   & 26.4            & 27.8            \\
\multicolumn{1}{l|}{deHuBERT (Ours)} & \multicolumn{1}{c|}{TEDLIUM}     & \textbf{19.3}   & \multicolumn{1}{c|}{\textbf{32.8}}   & \textbf{22.7}            & \textbf{22.8}            \\ \hline
\multicolumn{6}{c}{Testing set with additive OOD, office noise (0--20 dB)}                                                                        \\ \hline
\multicolumn{1}{l|}{HuBERT Base}     & \multicolumn{1}{c|}{LibriSpeech} & 26.6   & \multicolumn{1}{c|}{44.5}   & 42.2            & 43.9            \\
\multicolumn{1}{l|}{deHuBERT (Ours)} & \multicolumn{1}{c|}{LibriSpeech} & \textbf{17.0}   & \multicolumn{1}{c|}{\textbf{32.0}}   & \textbf{33.7}            & \textbf{35.5}            \\ \hline
\multicolumn{1}{l|}{HuBERT Base}     & \multicolumn{1}{c|}{TEDLIUM}     & 30.6   & \multicolumn{1}{c|}{46.2}   & 34.5            & 35.3            \\
\multicolumn{1}{l|}{deHuBERT (Ours)} & \multicolumn{1}{c|}{TEDLIUM}     &  \textbf{23.2}   & \multicolumn{1}{c|}{\textbf{37.4}}   & \textbf{26.2}            & \textbf{27.7}            \\ \hline
\end{tabular}
\label{tbl:2}
\end{table}

\section{Conclusion}
\vspace{-0.5mm}
In this paper, we proposed a novel pre-training framework that disentangles noise with the self- and cross-correlation loss for more robust speech recognition. Our model exhibits superiority in handling noisy ASR environments, including OOD noises, without compromising the performance of the clean audio test. The t-SNE plot of the contextual representations from deHuBERT offers a visual understanding of the improvement in noise robustness by observing randomly scattered projection that implies meagre embedded noise information.


\vfill\pagebreak

\bibliographystyle{IEEEbib}
\bibliography{strings,refs}

\end{document}